\documentclass[
 reprint,
 nofootinbib,
 amsmath,amssymb,
 aps,
 prb,
 pdfoutput=1,
]{revtex4-1}

\usepackage{graphicx}
\usepackage{xcolor}
\usepackage{bm}	
\usepackage{amssymb}
\usepackage{braket}
\usepackage{hyperref}
\usepackage{mathtools}
\usepackage{amsmath}
\usepackage{algorithm}
\usepackage{algpseudocode}
\usepackage{filecontents}

\newcommand{\da}{\downarrow}
\newcommand{\ua}{\uparrow}

\DeclareRobustCommand*{\citen}[1]{%
  \begingroup
    \romannumeral-`\x 
    \setcitestyle{numbers}%
    \cite{#1}%
  \endgroup   
}

\begin{document}
\title{Error retrodiction in quantum state tomography}
\title{Coarse-graining in retrodictive quantum state tomography}
\author{Dale Scerri}
\email{ds32@hw.ac.uk}
\affiliation{SUPA, Institute of Photonics and Quantum Sciences, Heriot-Watt University, David Brewster Building, Edinburgh, EH14 4AS, UK}
\author{Erik M. Gauger}
\affiliation{SUPA, Institute of Photonics and Quantum Sciences, Heriot-Watt University, David Brewster Building, Edinburgh, EH14 4AS, UK}
\author{George~C.~Knee}
\affiliation{Department of Physics, University of Warwick, Coventry, CV4 7AL, UK}

\date{\today}

\begin{abstract}
Quantum state tomography often operates in the highly idealised scenario of assuming perfect measurements. The errors implied by such an approach are entwined with other imperfections relating to the information processing protocol or application of interest. We consider the problem of retrodicting the quantum state of a system, existing prior to the application of random but known phase errors, allowing those errors to be separated and removed. The continuously random nature of the errors implies that there is only one click per measurement outcome -- a feature having a drastically adverse effect on data-processing times. We provide a thorough analysis of coarse-graining under various reconstruction algorithms, finding dramatic increases in speed for only modest sacrifices in fidelity. 
\end{abstract}

\maketitle
\section{Introduction}
Accurate quantum state reconstruction from finite data is a fundamental tool in quantum information science. Continued development of experimental tomography protocols and data-processing algorithms has improved both the accuracy and computational time required to produce state estimates in the face of the rapid increase in complexity of quantum systems. Despite being a mature field of research, quantum tomography -- covering state (QST), process (QPT) and detector tomography -- suffers from outstanding problems, such as state preparation and measurement (SPAM) errors. Whilst SPAM errors can be mitigated to some extent by using gate set tomography (GST) for gate characterisation, the latter is  significantly resource-intensive (requiring 4000 measurements to estimate a complete gate set, whereas only 256 are required to reconstruct a 2-qubit gate using QPT~\cite{Greenbaum2015}). In this work we shall deal with a particular type of SPAM error caused, for example, by noisy detector readout or by mis-calibrated measurement apparatuses. Measurement errors may be systematic or random, and will tend to reduce the fidelity of the tomogram, with respect to the true state $\rho$. If errors are known in a general quantum information processing protocol on a shot-by-shot basis, they may generally be compensated for by additional quantum control.  The irreversible nature of the quantum detection process, however, means that post-measurement knowledge of errors is insufficient for such compensation. 

\begin{figure}[t!]
\centering
\includegraphics[width=.8\linewidth]{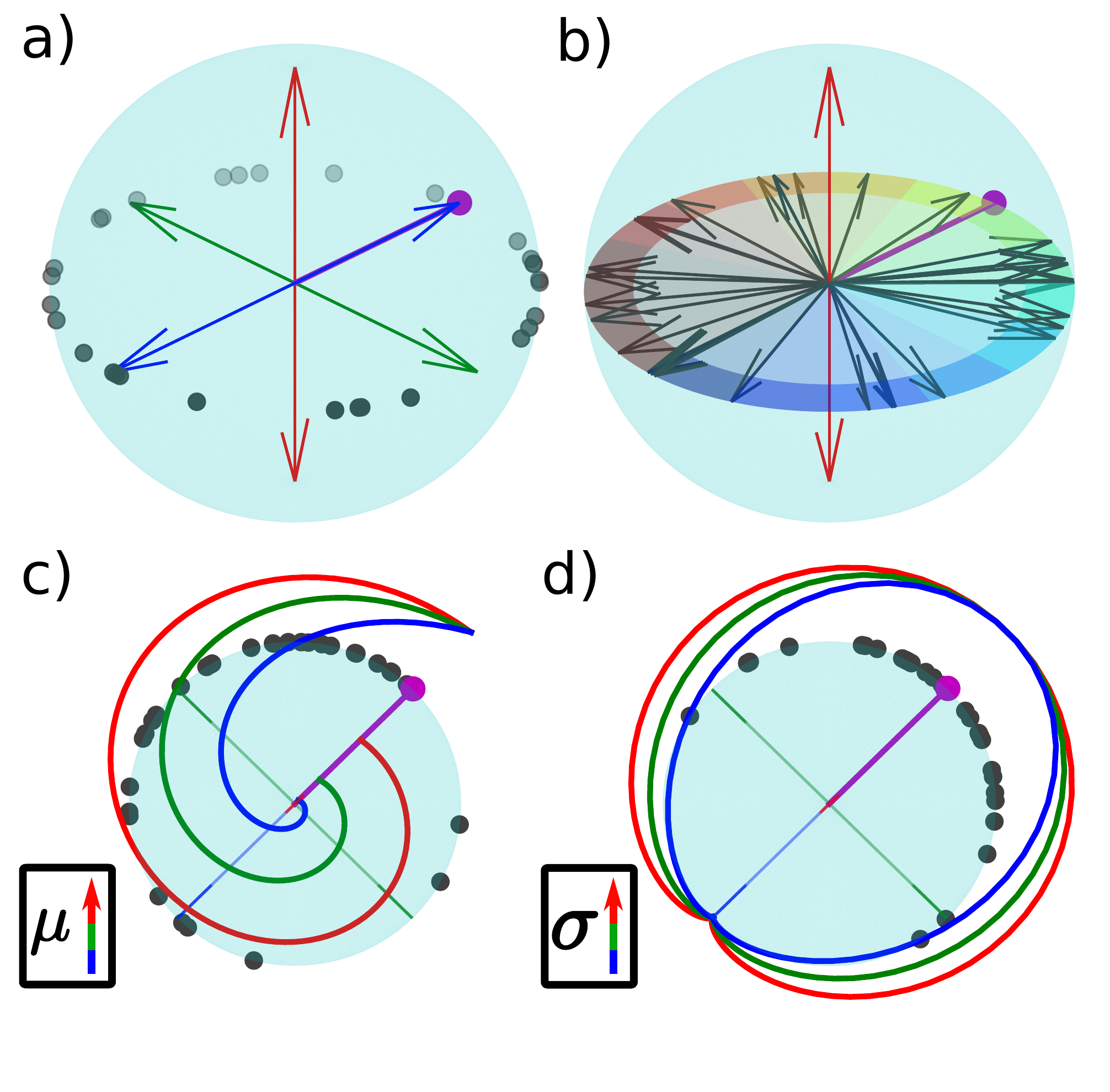} 
\caption{Bloch sphere representation of the problem in the context of a precessing qubit. \textbf{a)} In the Schr{\"o}dinger picture, the state (purple) gains a random phase (dots) prior to every measurement, with the measurement bases given by the arrows. \textbf{b)} In the Heisenberg picture, the state is static while the measurement operators are distributed randomly. The detector clicks can then be gathered in several bins on the Bloch sphere (coloured segments) to be used for coarse grained state reconstruction. \textbf{c)} Graphical depiction of exponentially distributed phases, for various means. \textbf{d)} Graphical depiction of normally distributed phases, for various distribution widths.}
\label{Bloch}
\end{figure}

Such a situation may be modelled by a semi-malevolent agent intervening in the experiment, applying random evolutions $\rho \rightarrow U_{\theta}\rho U^\dagger_\theta$ that are only revealed to the experimenter after they have made their measurements. For concreteness, we take $U_\theta = \cos\frac{\theta}{2}\mathbb{I}+i\sin\frac{\theta}{2}\sigma_z$ for $\sigma_z$ the usual Pauli operator, and $\rho$ as the system density matrix when no errors occur. Although the errors cannot be corrected in the sense of a fault tolerant quantum protocol, it is possible to retrodict the quantum state which existed before the errors were applied. Since the success probability of a fixed measurement operator $M$ is $p_\theta = \textrm{tr}(M[U_{\theta}\rho U^\dagger_\theta])= \textrm{tr}([U^\dagger_{\theta}MU_\theta]\rho)$, moving from the Schr\"odinger to Heisenberg pictures, the situation becomes equivalent to performing tomography on an ideal preparation $\rho$ with random measurements -- see Fig~\ref{Bloch}. The retrodiction is useful because $\rho$ may still contain other sources of error, which may then be separately estimated~\cite{Barnett2000a,Barnett2000b}.

A concrete example of such a situation comes from the field of photonic cluster state generation. A single emitter -- e.g. a natural atom or quantum dot -- will spontaneously undergo radiative decay at a random delay after excitation. The emitted photons are entangled with the emitter in such a way that repeated resonant control of the emitter's spin state and further excitations causes the subsequent emission of a chain of photons to be generated in a linear cluster state~\cite{Lindner2009, Scerri2018, Schwartz2016}: a key resource~\cite{Briegel2001} for measurement based quantum computation~\cite{Raussendorf2003}. Such schemes rely on an external magnetic field orthogonal to the optical axis \cite{Lindner2009, Scerri2018, Schwartz2016}. Due to the non-zero lifetime $\tau_{\textrm{decay}}$ of the emitter, the spin precesses at an angular frequency $\omega_l$ for a random interval. We may thus think of nature applying a random phase to the spin, which is then transferred to the emitted photon but revealed to the experimenter immediately upon detection. The task of estimating the density matrix $\rho$ of the photonic cluster state in the limit of $\tau_{\textrm{decay}}\rightarrow0$ is precisely the problem of retrodictive quantum state tomography outlined above.

For the technique to to work, it is necessary that the effective measurement operators are known:
In the precessing spin example, this information is revealed by the arrival time of the photon, the angular precession frequency $\omega_l$, and the time-of-flight of the photon to the detector. Because of the continuous nature of the distribution over $\theta$, the measurement record has the following `sparsity' feature: measurement operators will never be repeated, meaning that at most one click is attributed to each outcome. 
In this paper we show that retrodictive tomography is successful in spite of this feature, and go on to investigate the merits and demerits of coarse-graining -- a technique which removes sparsity by introducing a finite number of discrete bins which the measurement results are aggregated into. Our numerical simulations reveal that fidelity degrades monotonically as the number of bins is reduced, but that this is accompanied by a drastic improvement in algorithm run-time. As well as being a choice available to the tomographer, coarse-graining can also be considered as one way of simulating imperfect knowledge about the errors $\theta$. Intuitively, a Bayesian shot-by-shot approach is a natural paradigm to tackle the sparse tomography problem, making use of prior knowledge to process additional data obtained as more measurements are performed. However, the binning approach (discussed in Sec~\ref{sparse_and_binned}) cannot be applied to this technique straightforwardly. Thus, the Bayesian approach, as we shall see, suffers from being computationally expensive, but will still be be used as a benchmark for the Maximum Likelihood techniques which will follow.

In Section~\ref{sparse_and_binned}, we describe qualitatively how the sparse and coarse-grained QST methods work, outlining our methods for simulating tomographic datasets and assessing the performance of reconstruction algorithms. In Section~\ref{Bayesian} we introduce Bayesian estimation, along with an algorithm relying on a Monte Carlo implementation, followed by an outline of the Maximum Likelihood (ML) principle in Section~\ref{ML}, and an assessment of two distinct implementation algorithms. Section~\ref{norm_dist} treats normally distributed measurement operators, and we draw our conclusions in Section~\ref{conclusions}. Finally, we give the full details of the algorithms used, along with some additional results, in the Appendix.

\section{Sparse and binned tomography}
\label{sparse_and_binned}

The distribution $p(\theta)$ (supported on $[0, 2\pi)$) of effective measurement operators depends on the physical scenario: in the example of frequency cluster-state generation in the hole-spin system in Ref.~\citen{Scerri2018}, when the precession time is much shorter than the emission time, $p(\theta) \approx 1/(2\pi)$. In such a case, the coherences of the reconstructed state would be completely washed out by conventional QST techniques (not making use of the knowledge of the errors $\theta$). For the more general case of photon emission from spin-bearing emitters, however, the exponential distribution  $p(\theta) \propto \mathrm{e}^{- \theta / \mu}$ (with the mean $\mu = \lambda^{-1}$, where $\lambda$ is the rate parameter) is more adequate to describe the spread of operators. Other distributions may be similarly treated -- meaning that our analysis applies to a wider range of physical scenarios --  although the measurement operators may then be clustered to a greater or lesser degree, having an affect on the accuracy of the retrodicted tomogram. The normal distribution $p(\theta)\propto \mathrm{e}^{-\theta^2/2\sigma^2}$ ($\sigma$ being the standard deviation) is considered in Section~\ref{norm_dist}, while as $\mu \rightarrow \infty$, we recover the uniform distribution limit, i.e. $p(\theta) \rightarrow 1/(2 \pi)$.

In the Schr\"odinger picture, we fix the four measurement operators $\Ket{\ua}\Bra{\ua}$, $\Ket{\da}\Bra{\da}$ and $\Ket{\phi}\Bra{\phi}$, where $\sqrt{2}|\phi\rangle = |\uparrow\rangle+e^{i\phi}|\downarrow\rangle$ and $\phi\in\{0,\pi\}$. Since emitted photons are measured independently, $m$-qubit states are tomographed by forming $m$-fold tensor products of all combinations of these projectors. 

By using the Heisenberg picture (as in the previous section), the tomographic protocol is equivalent to reconstructing some unknown state $\rho$ with the following set of positive (projective)  measurement operators
 \begin{align}\label{basis}
\mathcal{P} = \{ |\uparrow\rangle\langle \uparrow| ,|\downarrow\rangle\langle \downarrow|,M_{\theta_i} = U^\dagger_{\theta_i} |\phi\rangle\langle \phi|U_{\theta_i}\}
\end{align}
where $U^\dagger_{\theta_i} |\phi\rangle=|\phi+\theta_i\rangle$ for $\theta_i$ ($i=1,\ldots,N$) drawn from $p(\theta)$. Note that $ U^\dagger_{\theta_i}|\updownarrow\rangle \langle \updownarrow | U_{\theta_i}= |\updownarrow\rangle \langle \updownarrow |$, and that the values of $\phi$ play less of a role as the spread of $\theta$ increases. Because $\Ket{\phi}\Bra{\phi} + \Ket{\phi + \pi}\Bra{\phi + \pi} = \mathbb{I}$, this set may be considered a POVM (Positive Operator Valued Measure) upon appropriate normalisation (in the sense that the sum of all operators is proportional to the identity). 

We generated pseudo-tomographic data for a fixed $\rho$ by drawing $N/2$ unique values of $\theta\in[0,2\pi)$ from $p(\theta)$. We then simulate a single Bernoulli trial for each measurement operator, assigning the event to $M_{\theta_i}$ with probability $p_i=\textrm{tr}(\rho M_{\theta_i})$, and to the orthogonal operator with the complementary probability. The measurement record then consists of a (multi)set of $N/2+2$ measurement operators with (for the $N/2$ operators perpendicular to the `precession' axis) multiplicities $n_i=1$, and the two orthogonal operators (parallel to the `precession' axis) with joint multiplicity of $N/2$ (i.e. it is `sparse'). For the former, $N/2$ measurements are then split between the two projections along the precession axis. Optionally, we modify the measurement record by a process of coarse-graining or `binning', resulting in a lower number $N_b<N/2$ of coarse-grained measurement operators, e.g.
\begin{equation}
\tilde{M}_{\theta_j}=|2\pi/N_b\rangle\langle 2\pi/N_b|~,
\end{equation}
projecting onto states evenly distributed around the equator of the Bloch sphere (see Fig.~\ref{Bloch}) with multiplicities
\begin{equation}
\tilde{n}_j = \sum_i n_i \textrm{rect}\left(\frac{N(\theta_i -\theta_j)}{2\pi}\right)~,
\end{equation}
that simply accumulate the events according to the bin that they fall within (with the bins being intervals centred on $\theta_j$ and with width $N/2\pi$, as shown graphically in Fig.~\ref{Bloch}). Other binning schemes are possible, including those that depend on the original measurement record~\cite{Glancy2018}. We then run different reconstruction algorithms (to be introduced below) on the coarse-grained measurement record, to give a quantum state estimate or `tomogram' $\rho_{est}$. The running time of the algorithm is noted, and the fidelity of the tomogram computed: $F(\rho_{est},\rho)=\textrm{tr}\sqrt{\sqrt{\rho}\rho_{est}\sqrt{\rho}}$. The infidelity is $1-F$, and is a measure of the distance between the true state and the retrodicted tomogram. The procedure was then repeated for distinct, randomly generated (but full rank) $\rho$, and we collected statistics to summarise the typical performance.

Counter to intuition, using sparse tomography without any binning works remarkably well. However, algorithm running time tends to scale badly with $N$ (since the calculation of the cost function and its gradient involves a contribution from each of the $N$ distinct operators). Hence our proposed coarse-grained approach. The remainder of the paper is dedicated to investigating the dependence of fidelity and run time on $N_b$, for different reconstruction algorithms. As $N_b,N\rightarrow \infty$, the sparse and coarse grained approaches are expected to give the same fidelities.

\section{Non-adaptive Bayesian tomography}
\label{Bayesian}

The Bayesian approach was introduced in the field of quantum tomography \cite{Jones1991a, Jones1991b, Slater1995, Derka1997, Buzek1998, Schack2001}, and is an ongoing theoretical and experimental research topic \cite{Kohout2010, Granade2016, Struchalin2016}. This approach offers numerous advantages over other techniques, such as use of online information available to the experimentalist after each measurement. Furthermore, Bayesian inference was also shown to be optimal with respect to any strictly proper scoring rule derived from Bregman distances \cite{Kohout2006,Kohout2010,Granade2012} (near-optimal if the infidelity is used as a loss function instead \cite{Kueng2015}), with the ability to track fidelity bounds online \cite{Kueng2015} (allowing for feedback to minimise number of required measurements), as well as giving robust region estimates \cite{Ferrie2014} and allowing for model selection/averaging. Thus the Bayesian approach shall be used as a benchmark for the other techniques discussed in this work.

Our implementation follows closely the approaches used in Refs. \citen{Granade2016} and \citen{Huszar2012}. For a Bayesian update scheme, we start with an initial prior probability density $p(\rho)$ over feasible state space (usually uninformed due to the absence of additional knowledge, resulting in a uniform prior). After obtaining a new measurement datum $D$, the posterior distribution $p(\rho| D)$ is then built using the likelihood function $\mathcal{L}(\rho;D)$ as 
\begin{equation}\label{posterior}
p(\rho| D) \propto \mathcal{L}(\rho;D) p(\rho)~.
\end{equation}
Typically, Bayesian tomography schemes would then make use of the narrower posterior and additional criteria (for example, Shannon information \cite{Huszar2012}) to infer the next optimal measurement setting \cite{Huszar2012,Granade2016}. However, since we do not have control over which measurement to perform next, this latter step of the Bayesian scheme  cannot be applied. Although we do not make use of any criteria to track the narrowing of the sample, one could still use the covariance of the the narrowed posterior, in this case, to indicate when a sufficiently precise estimate has been found.

\begin{figure}[t!]
\includegraphics[width=1\linewidth]{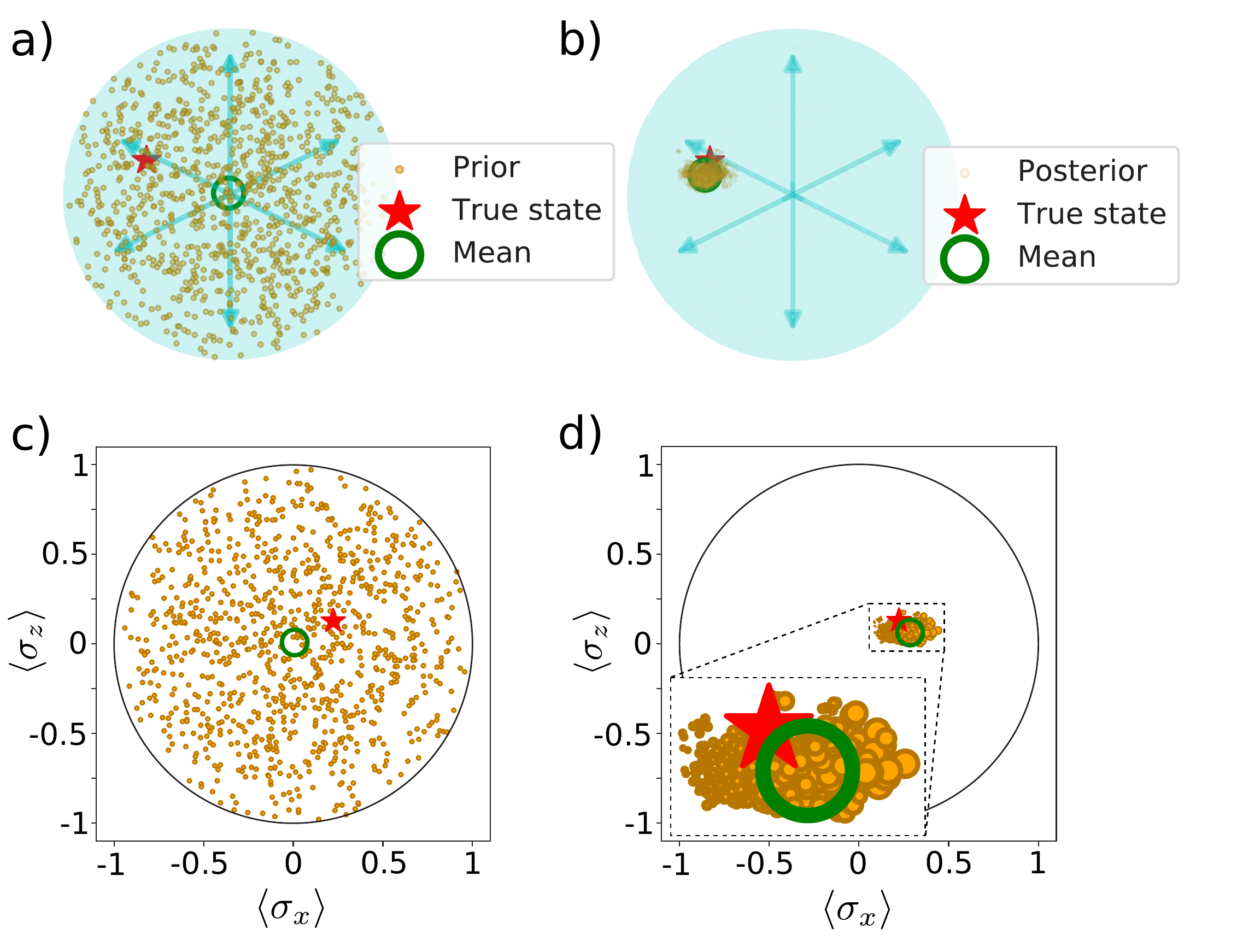}
\caption{\textbf{a)} Initial uninformed prior (orange), with the mean of the distribution shown in green, and the true state to be reconstructed in red. \textbf{b)} Final posterior (orange) after 2000 measurements, where the marker size indicates the relative particle weights. \textbf{c)} and \textbf{d)} show the $[\langle \sigma_x \rangle, \langle \sigma_z \rangle]$ projection of the prior and posterior, respectively, as a visual aid. As more measurements are performed, most of the original particle weights drop to zero, requiring resampling for a more accurate prediction without requiring an excessive number of particles to begin with.}
\label{bay}
\end{figure}
\footnotetext{}
\begin{figure}[t!]
\center
\includegraphics[width=.9\linewidth]{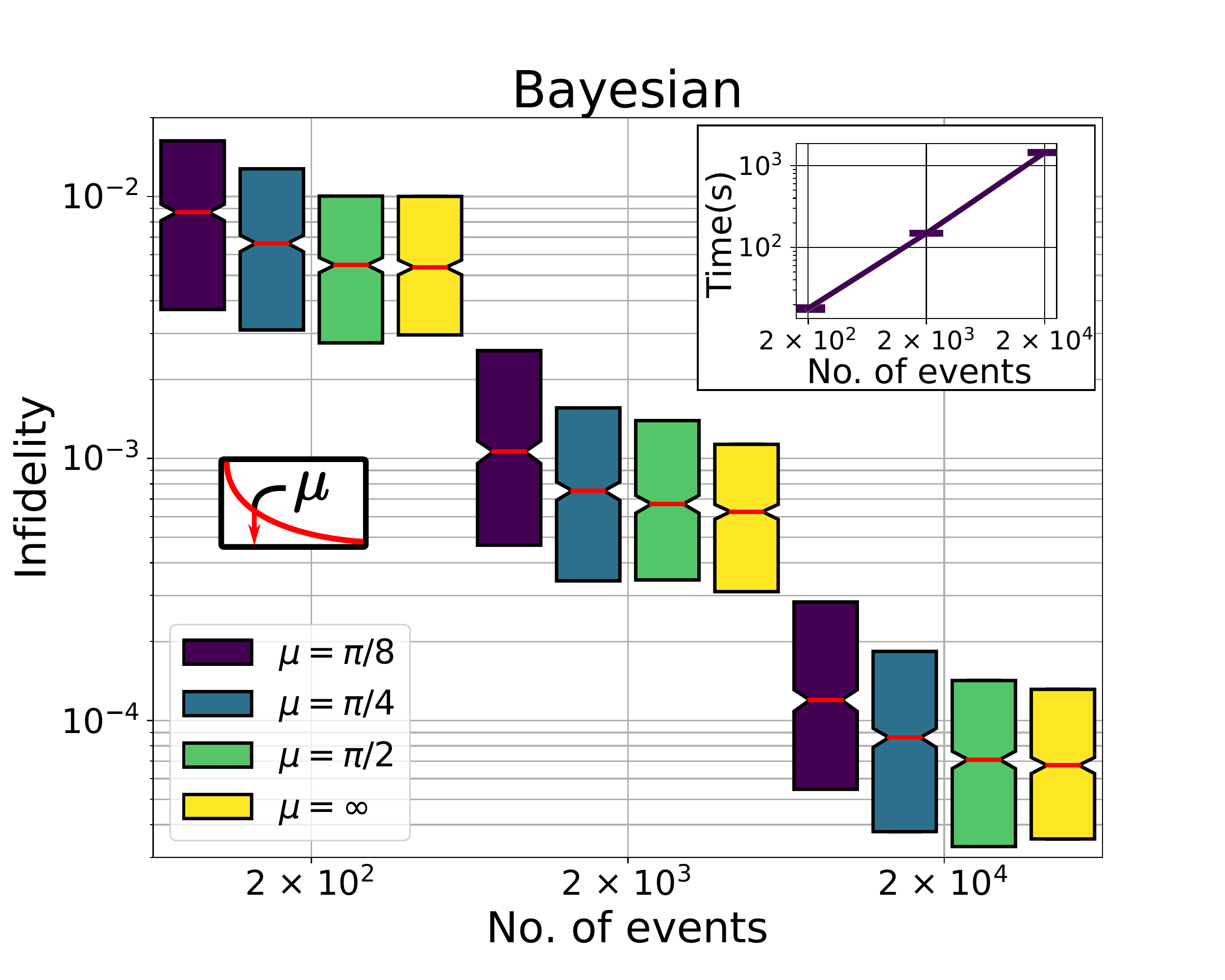}
\caption{First and third quartile box plots for full rank, single qubit reconstruction using the non-adaptive Bayesian approach, with exponentially distributed measurement operators, averaged over 1000 trials using 1000 particles for sampling. The performance improves with the rate parameter $\mu$. Unless otherwise stated, all box plot error bars will display first and third quartiles. Otherwise, error bars used correspond to one sigma uncertainty. \textbf{Inset:} Algorithm running times for the Bayesian approach. For all values of $\mu$, the computation time scales linearly with number of measurement repetitions (or, equivalently, the number of operators used for the reconstruction) due to the sparse nature of our Bayesian reconstruction.
}
\label{bay_exp}
\end{figure}
Despite the simple form of Eq.~\ref{posterior}, the analytical evaluation of the posterior is seldom feasible, and hence the latter is typically replaced with an approximation. To this end, several Markov Chain Monte Carlo techniques (MCMC)  have been adopted, including the Metropolis-Hastings algorithm \cite{Kohout2010}. However, these MCMC techniques tend to be computationally expensive, with decreasing acceptance probabilities at each sampling step, leading to more samples being discarded as additional data is obtained. Furthermore, these methods require the assumption of a normal posterior, which is not always the case in state tomography. The Sequential Monte Carlo technique (SMC) \cite{Liu1998, Liu2001}, on the other hand, only requires the computation of a single term of the likelihood to update the weights of the approximate distribution with each measurement \cite{Huszar2012}. In this approach, adopting the notation in Ref.~\citen{Huszar2012}, the posterior after the $i^\mathrm{th}$ measurement is approximated by a number $P$ of randomly sampled particles, $\{\rho_p\}$, and their corresponding weights $\{w^{(i)}_p\}$ as
\begin{equation}
p(\rho|\{D_i\}) \approx \sum^P_{p=1} w^{(i)}_p \delta(\rho - \rho_p).
\end{equation}

Suppose our current (prior) knowledge is given by the dataset $\{D_i\} = \{ \alpha_j:~ 1 \leq j \leq i,~ \alpha_j \in \mathcal{P} \}$, where the set $\mathcal{P}$ is defined in \eqref{basis}. If the next projection phase is, without loss of generality, $\theta_{i+1}$, (that is, $\alpha_{i+1} = M_{\theta_{i+1}}$), then, following Ref.~\citen{Huszar2012} and using Bayes' rule (Eq.~\ref{posterior}), we can write the approximation for the next posterior as

\begin{align}
\begin{split}
p(\rho|\{D_{i+1}\}) &= p(\rho|\{D_i\} \cup \{M_{\theta_{i+1}}\}) \\
&\approx \sum^P_{p=1} \frac{\mathbb{P}(M_{\theta_{i+1}}|\rho_p) w^{(i)}_p}{\sum^P_{q=1} \mathbb{P}(M_{\theta_{i+1}}|\rho_q) w^{(i)}_q} \delta(\rho - \rho_p) \\
&\coloneqq \sum^P_{p=1} w^{(i+1)}_p \delta(\rho - \rho_p)~,
\end{split}
\end{align}
where $\mathbb{P}(M_{\theta_{i+1}}|\rho_p) = \mathrm{Tr}(M_{\theta_{i+1}}\rho_p)$. In our numerical simulations, we do the first $N/2$ measurements along the z-axis (that is, using projection operators $\{ \Ket{\ua}\Bra{\ua}, \Ket{\da}\Bra{\da} \}$), followed by the remaining $N/2$ measurements along the Bloch equatorial plane. As more measurements are performed, narrowing the particle distribution, most of the weights drop to zero, which can be remedied by resampling using the new posterior distribution \cite{Huszar2012}. Finally, the Bayes estimator $\rho_{est}$ can be extracted from the mean of the final posterior approximation. In Fig.~\ref{bay} we show the above steps graphically, emphasising the use of resampling to obtain an accurate posterior.

We numerically benchmarked the Bayesian technique, using a uniform prior \footnote{Samples are drawn from a Ginibre ensemble (random matrices with normally distributed entries), and subsequently squared to give positive matrices. Finally, these are normalised to unit trace \cite{Johansson2012, Johansson2013}}. An example is shown in Fig.~\ref{bay}. and further results are summarised in Fig.~\ref{bay_exp}. Despite the fact that we cannot decide which measurement to perform next, our random basis measurement can be seen to give a good convergence after 2000 measurements with 1000 particles. 

\section{Maximum Likelihood Estimation}
\label{ML}
A common, alternative, approach to state estimation is producing a tomogram $\rho_{est}$ which maximises the likelihood function. Naive approaches may result in an invalid tomogram (having, for example, negative eigenvalues). The search for the best fit to the data, therefore, should be constrained to the allowed state space of trace-one positive semidefinite matrices~\cite{Kaznady2009, Teo2016, Knee2018}). Previously (in the Bayesian method) this was ensured by choosing a prior distribution supported only in the allowed state space. Here, the prior is not modelled, but we consider two alternative approaches: A) the constraints are enforced by a non-linear parametrization of the density matrix and B) the constraints are enforced periodically in the course of an iterative gradient descent procedure, allowing for temporary violations \cite{Goncalves2016, Bolduc2017, Knee2018}. Given a density matrix $\rho$, the likelihood function to be maximised has the form
\begin{equation}
\mathcal{L}(\rho) = \prod^{N_b}_{j=1} p_j^{n_j}~,
\end{equation}
with equality holding up to an irrelevant proportionality constant. For sparse tomography, the product would be over $N$ exponentiated probabilities $p_j$, with each $n_j$ taking a binary value of either 0 or 1. Due to the monotonicity of the logarithm, maximising the likelihood function is identical to minimising the negative of its logarithm [which we refer to as the cost function $\mathcal{C}(\rho)$], given by
\begin{equation}\label{negloglik}
\mathcal{C}(\rho) \coloneqq -\mathrm{log}~\mathcal{L}(\rho) = -\sum^{N_b}_{j=1} n_j \mathrm{log}(p_j)~,
\end{equation}
where we took the normalising constant to identity. Recall that the sparse tomography limit is recovered when $N_b=N$ and $n_j=1$. In the limit of a large number of detections per measurement, the probability of obtaining the $j^\mathrm{th}$ measurement can be approximated by a Gaussian distribution \cite{James2001, Altepeter2005}, with the estimated number of detections for the $j^\mathrm{th}$ measurement given by $ \bar{n}_j = N p_j$. Since this approximation clearly fails for the sparse case due to the binary nature of the $n_j$'s, we do not make it.

\subsection{Cholesky factorisation}

\begin{figure}[t!]
\centering
\includegraphics[width=\linewidth]{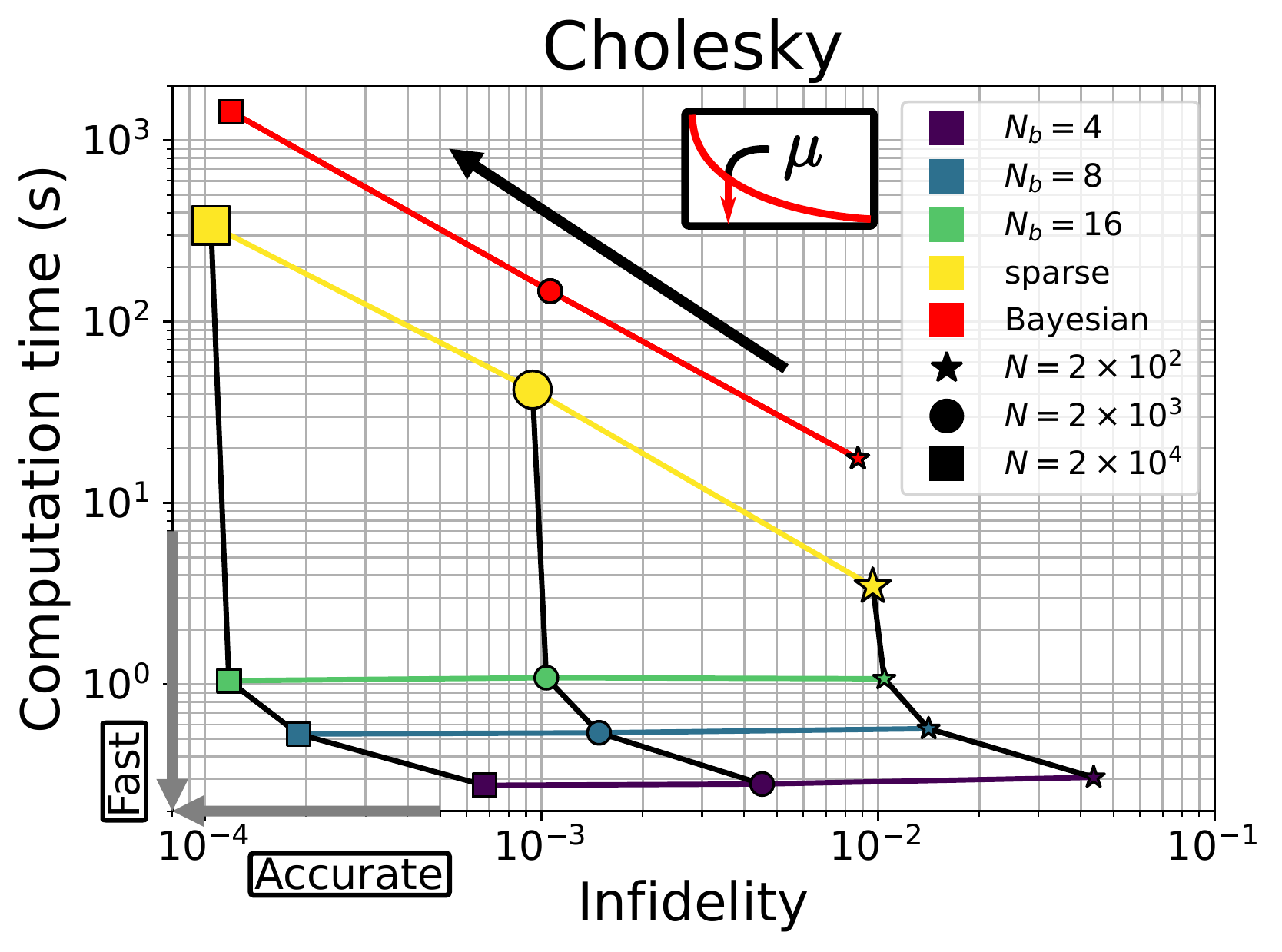} 
\caption{Full rank, single qubit reconstruction using the Cholesky decomposition method, averaged over 1000 trials. The random phases were sampled from an exponential distribution with  $\mu = \pi/8$. As expected the coarse grained approach returns slightly higher infidelities (shown on the x-axis). The algorithm running times (y-axis) for the sparse approach scales linearly with number of measurement repetitions. On the other hand, the computation times for the binned approach, within error bars, remain the same with increased repetitions, as the number of projective operators used for reconstruction is the same for all repetition numbers. The results from the Bayesian method are also shown for comparison. While the Bayesian approach offers higher fidelity estimates for lower measurement numbers $N$ (star), the infidelity is higher compared to the sparse PGDB for higher $N$, and the corresponding computation time heavily offsets any advantages gained in fidelity by the Bayesian approach. The black arrow indicates the direction of the trend as the number of measurement events increases; infidelity decreasing at the expense of higher computation time, whilst the grey arrows on the axes point towards the ideal region of low infidelity and computation time.}
\label{1Q}
\end{figure}
In this section we implement a Cholesky-like decomposition of the density matrix in order to minimise Eq.~\ref{negloglik} \cite{James2001, Altepeter2005, Goncalves2011}, allowing us to use Python's SciPy least-squares solver on a 1D array \footnote{More formally, this least-squares solver uses the Trust Region Reflective technique (involving searching along directions reflected from the trust region bounds).} One can easily show that any qubit density matrix $\rho$ allows for a decomposition of the form
\begin{figure*}[t!]
\centering
\includegraphics[width=\linewidth]{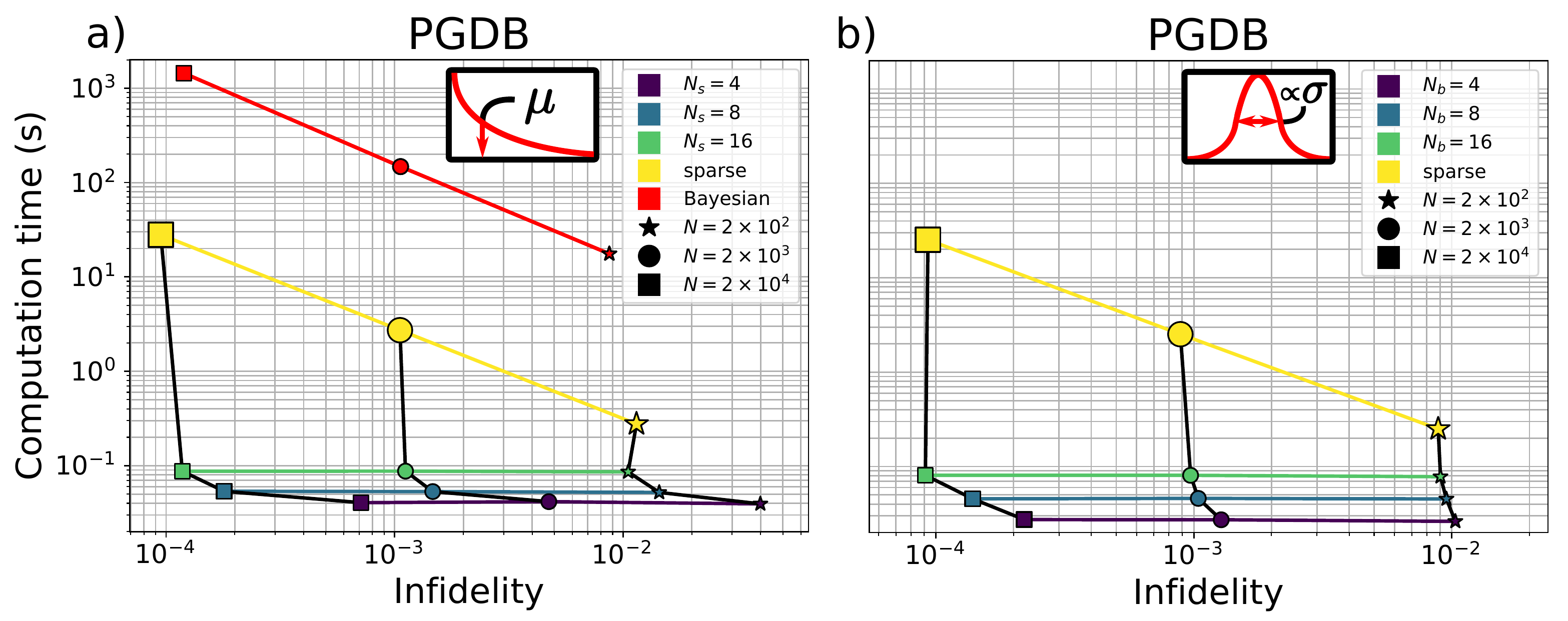} 
\caption{Full rank, single qubit reconstruction using gradient descent, averaged over 1000 trials. \textbf{a)} The random phases were sampled from an exponential distribution with $\mu = \pi/8$. The results follow a similar behaviour as the Cholesky method, except that the PGDB algorithm (for the given exit criteria in Appendix~\ref{app:pseudo}) shows lower computation times both for the sorted and binned approaches. For comparison, we also show the Bayesian result for the exponentially distributed phases. \textbf{b)} The random phases were sampled from a normal distribution with standard deviation $\sigma = \pi/8$. For lower $N$, going from $N_b = 4$ to $N_b = 8$ or from $N_b = 16$ to sparse tomography does not reduce the infidelity as significantly as when increasing the number of bins from 8 to 16.}
\label{pgdb_coarse_alt}
\end{figure*}
\begin{equation}
\rho = T^\dagger T / \mathrm{Tr}[T^\dagger T]~,
\end{equation}
where $T$ is the lower triangular matrix given by
\begin{equation}
T(\mathbf{t}) = \left(\begin{array}{cc}t_1 & 0 \\t_3 + i t_4 & t_2\end{array}\right)~,
\end{equation}
with $\mathbf{t} = (t_1,t_2,t_3,t_4)$ being the array over which the minimisation search is performed. In particular, we can use this decomposition to calculate $\bar{n}_j \propto p_j = \mathrm{Tr}\left[ \Ket{\phi+\theta_j}\Bra{\phi+\theta_j} T^\dagger T \right]/ \mathrm{Tr}[T^\dagger T] $. Generalising this parametrisation to $m$ qubits, we get
\begin{equation}
T(\mathbf{t}) = \left(
\begin{array}{cccc}t_1 & 0 & ... & 0 \\
t_{2^m +1} + i t_{2^m +2} & t_2 & ... & 0 \\
... & ... & ... & 0 \\
t_{4^m -1} + i t_{4^m} & t_{4^m -3} + i t_{4^m-2} & ... & t_{2^m}\end{array}\right)~,
\end{equation}
and hence the search needs to be done over a real array of length $4^m$. 

Having formulated a decomposition guaranteeing a valid density matrix, the problem can be recast to a least-squares minimisation problem \cite{James2001, Altepeter2005} in order to find the minimum of the negative log likelihood, as the latter may be written down as
\begin{equation}\label{loglik2}
\mathcal{C}(\rho)  = \sum^N_{i=1} [f_i(\mathbf{t})]^2~,
\end{equation}
where, for the general case of a multinomial probability distribution, we get using Eq.~\eqref{negloglik}
\begin{align}
\begin{split}
f_j(\mathbf{t}) =  &\sqrt{n_j \mathrm{log}(p_j)} \\
= &\sqrt{n_j} \left(\mathrm{log}\{\mathrm{Tr}\left[ \Ket{\phi_j} \Bra{\phi_j} T^\dagger(\mathbf{t})T(\mathbf{t}) \right] \}\right. \\
&\left.\hspace{8mm}- \mathrm{log}\{ \mathrm{Tr}\left[T^\dagger(\mathbf{t})T(\mathbf{t})\right] \} \right)^{\frac{1}{2}}~.
\end{split}
\end{align}
Despite having multiple local minima, this optimization problem was shown to have a single global solution \cite{Goncalves2011}, meaning that all local minimizers lead to the same solution minimizing the negative log likelihood.

In Fig.~\ref{1Q} we show the results for single qubit reconstruction. As expected, the fidelity of the reconstructed density matrix increases with number of Bloch sphere partitions. This is also the case for a two-qubit reconstruction, as we show in Appendix~\ref{app:TwoQubit}.

\subsection{Projected gradient descent}

Gradient descent algorithms rely on following the path of steepest descent of the cost function, in this case Eq. \eqref{negloglik}, starting from a well chosen initial estimate. If left unconstrained in the convex space of $d \times d$ matrices (where $d$ is the Hilbert space dimension), the resulting estimate $\rho_{est}$ might lie outside the convex subspace of unit-trace, positive semidefinite matrices, leading to an unphysical estimate. Hence, projection back to the physical subspace, minimising distance as measured through of a matrix norm (such as projection of the spectrum onto the unit simplex \cite{Goncalves2011, Goncalves2016, Bolduc2017}) is employed, giving rise to projected gradient descent (PGD) algorithms. Iterating this process leads to a convergence of the cost function to a minimum below a predefined threshold. A unique solution satisfying the appropriate constraints and minimising the cost function is then guaranteed as long as the latter is a continuously differentiable convex function of the density matrix. Eq.~\eqref{negloglik} is convex but not continuously differentiable, but this tends to not pose a problem in practice, as discussed in Ref.~\citen{Knee2018}. Choosing the projection of $\rho$ to be of its spectrum onto the unit simplex (which we refer to as $\mathcal{P}_S$), the PGD algorithm update can be written as
\begin{equation}\label{pgd}
\rho_k = \mathcal{P}_S[\rho_{k-1} - \nabla \mathcal{C}(\rho_{k-1})]~.
\end{equation}

As is commonplace, we supplement the PGD algorithm with a backtracking line search (PGDB) based on the Armijo--Goldstein condition to losely optimise the maximum step size for each descent iteration \cite{Goncalves2011, Goncalves2016, Bolduc2017}. The estimate at the $k^\mathrm{th}$ PGDB iteration can thus be written as
\begin{equation}\label{pgdb}
\rho_k = (1-\alpha)\rho_{k-1} + \alpha~\mathcal{P}_S[\rho_{k-1} - \nabla \mathcal{C}(\rho_{k-1})]~,
\end{equation}
where $\alpha \in \left[0,1\right]$ is the line search parameter to be roughly optimised at each step. We assess the impact of binning on the PGDB algorithm, Fig.~\ref{pgdb_coarse_alt}a showing the trade-off between computation time and fidelity for $\mu = \pi/8$. Fig.~\ref{pgdb_coarse_alt}b, on the other hand shows the relation between computation time and infidelity for various number of bins $N_b$ and events $N$ for normally distributed phases, showing a similar trend to the exponentially spread phases. Appendix~\ref{app:PGDBalt} shows a closer analysis of the exponential data, with first and third quartiles for the infidelity, and standard deviation errorbars for computation times.

\begin{figure}[t!]
\center
  \includegraphics[width=.95\linewidth]{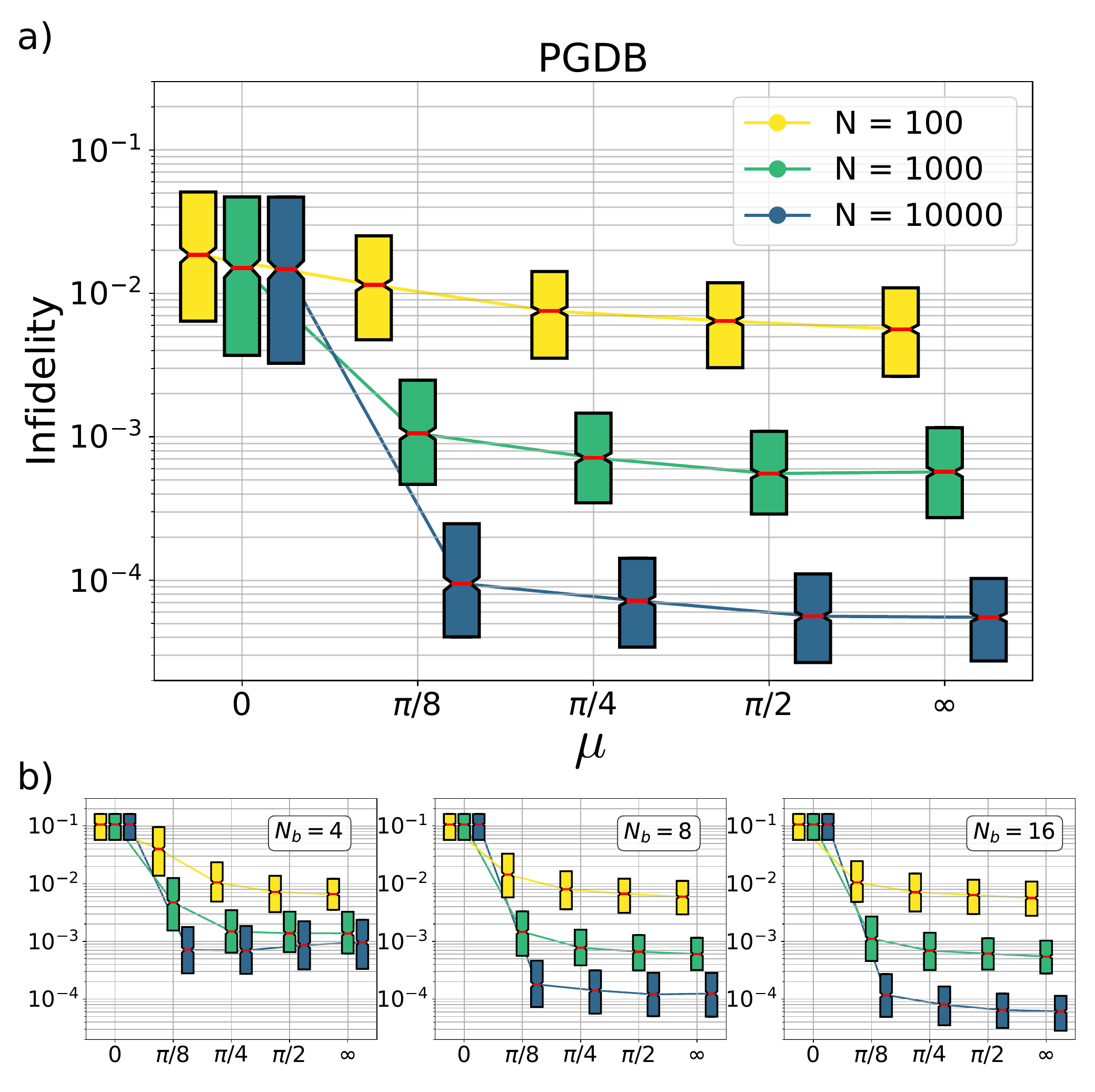}
\caption{\textbf{a)} Full rank, sparse single qubit reconstruction infidelities for phases sampled from exponential distribution with various values of $\mu$ and experiment repetitions $N$. \textbf{b)} Infidelities for various segment numbers $N_b$. In both \textbf{a)} and \textbf{b)}, averages were performed over 1000 trials.}
\label{ExpMulti}
\end{figure}

As expected, within standard deviation error, the binned approach gives slightly lower fidelities than the sparse one. This difference, however, is well justified when considering the significant reduction in computation time shown in Fig.~\ref{pgdb_coarse_alt}a. The trends in Fig.~\ref{pgdb_coarse_alt}a, both for computation time and infidelity, are similar to those shown in Fig.~\ref{1Q} for the Cholesky method. However, our numerical simulations clearly show lower reconstruction times achieved using the PDGB technique. In Fig.~\ref{ExpMulti}, we show how the infidelity varies with increasing mean $\mu$ for various values of the bin number $N_b$.

\section{Condition numbers}
\label{norm_dist}

\begin{figure}[t!]
\includegraphics[width=.9\linewidth]{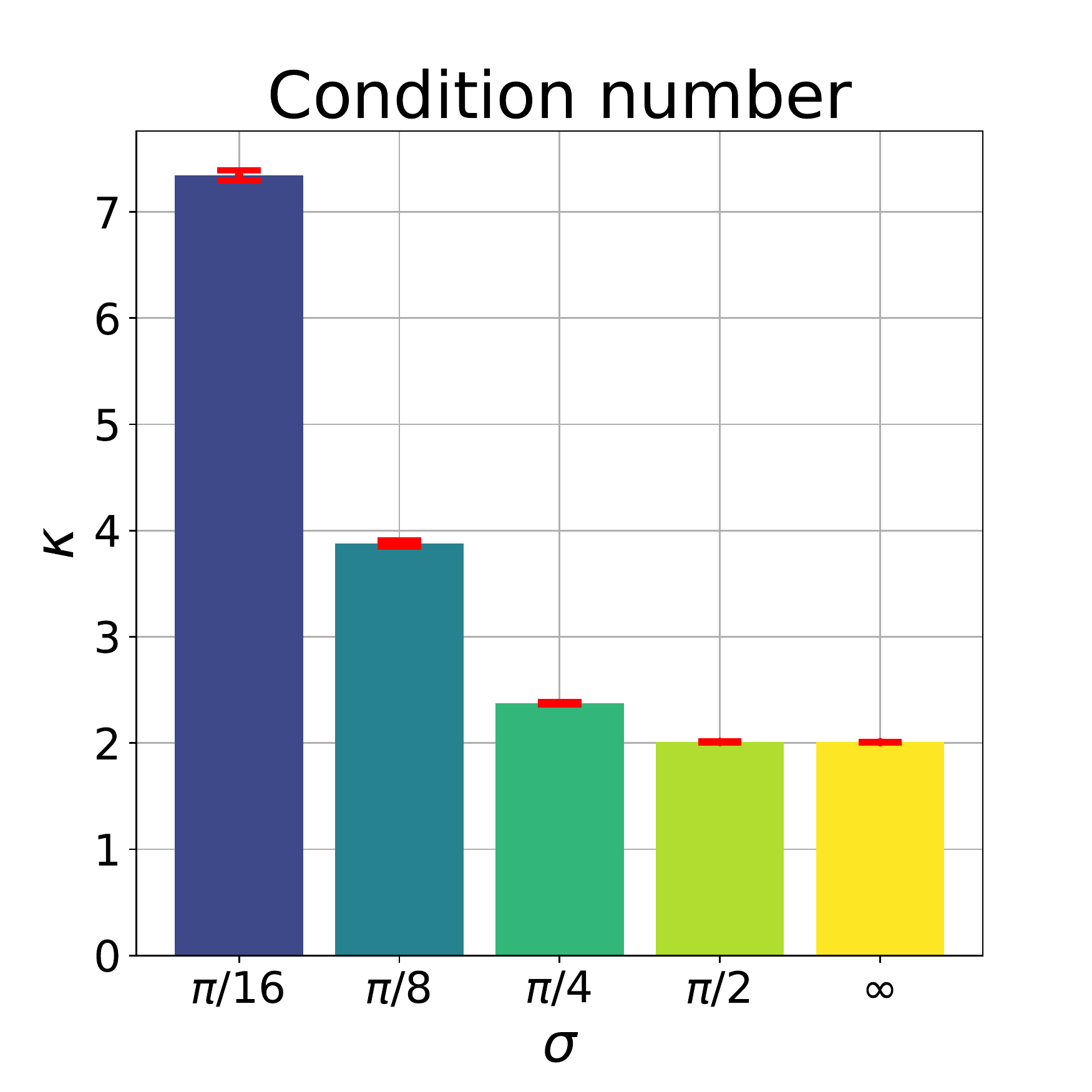}
\caption{Sparse tomography condition number (shown above for $N=2\times 10^4$) decreases (improves) as the standard deviation of the normally distributed phases ($\sigma$) increases. The red bars indicate one sigma uncertainty. When $\sigma$ is high, we recover the limit of many measurements distributed evenly around the equator of the Bloch sphere. In this situation, we obtain the same condition number, $\kappa(A)=2$, as in the case of complete Pauli measurements \cite{Miranowicz2014}.}
\label{cond}
\end{figure}

Using a single basis for reconstruction along the plane of precession, we see that the higher the spread of the distribution, the higher the fidelity one expects, as the effective rotated bases sample larger portions of the Bloch plane, whereas for lower spreads, the additional phase knowledge does not contribute considerably, and hence incomplete Pauli tomography (in which only x- and z- basis measurements are performed) is recovered. This can be seen in Fig.~\ref{cond}, showing the behaviour of the condition number $\kappa(A)$ of the measurement matrix $A$ for increasing $N$, where $A$ is given by 
\begin{equation}
A = \left( \begin{array}{c}
 \mathrm{vec}(\hat{\Pi}_1)^T  \\ 
 \vdots \\ 
 \mathrm{vec}(\hat{\Pi}_{N+2})^T \\ 
 \end{array}\right) ~,
 \end{equation}
where the projectors $\hat{\Pi}_i$ make up the set $\mathcal{P}$ in Eq.~\eqref{basis} \cite{Miranowicz2014, Bolduc2017}. The condition number decreases significantly with increasing standard deviation of the distribution, meaning that sampling distributions with larger spreads results in a better conditioned measurement matrix.  

\section{Conclusion}
\label{conclusions}
QST is still an active area of experimental and theoretical research, allowing the reconstruction of quantum states from finite experimental data. In this work, we implemented several QST algorithms in the presence of phase errors which is only known after the system is measured. We showed, with a simple modification, how the unaffected state may be retrodicted using such knowledge. Furthermore, we demonstrated that, at a small cost in fidelity, the reconstruction time can be significantly decreased. All data in this work was generated and visualised using Python and QuTiP package \cite{Johansson2012, Johansson2013}.

\begin{acknowledgments}

We thank Cristian Bonato for insightful and stimulating discussions. D.S. thanks SUPA for financial support. G.C.K. was supported by the Royal Commission for the Exhibition of 1851, and E.M.G. acknowledges support from the Royal Society of Edinburgh and the Scottish Government.

\end{acknowledgments}

\appendix
\setcounter{figure}{0} \renewcommand{\thefigure}{A.\arabic{figure}} 

\section{Two-qubit results}
\label{app:TwoQubit}

\begin{figure}[h!]
\center
\includegraphics[width=.7\linewidth]{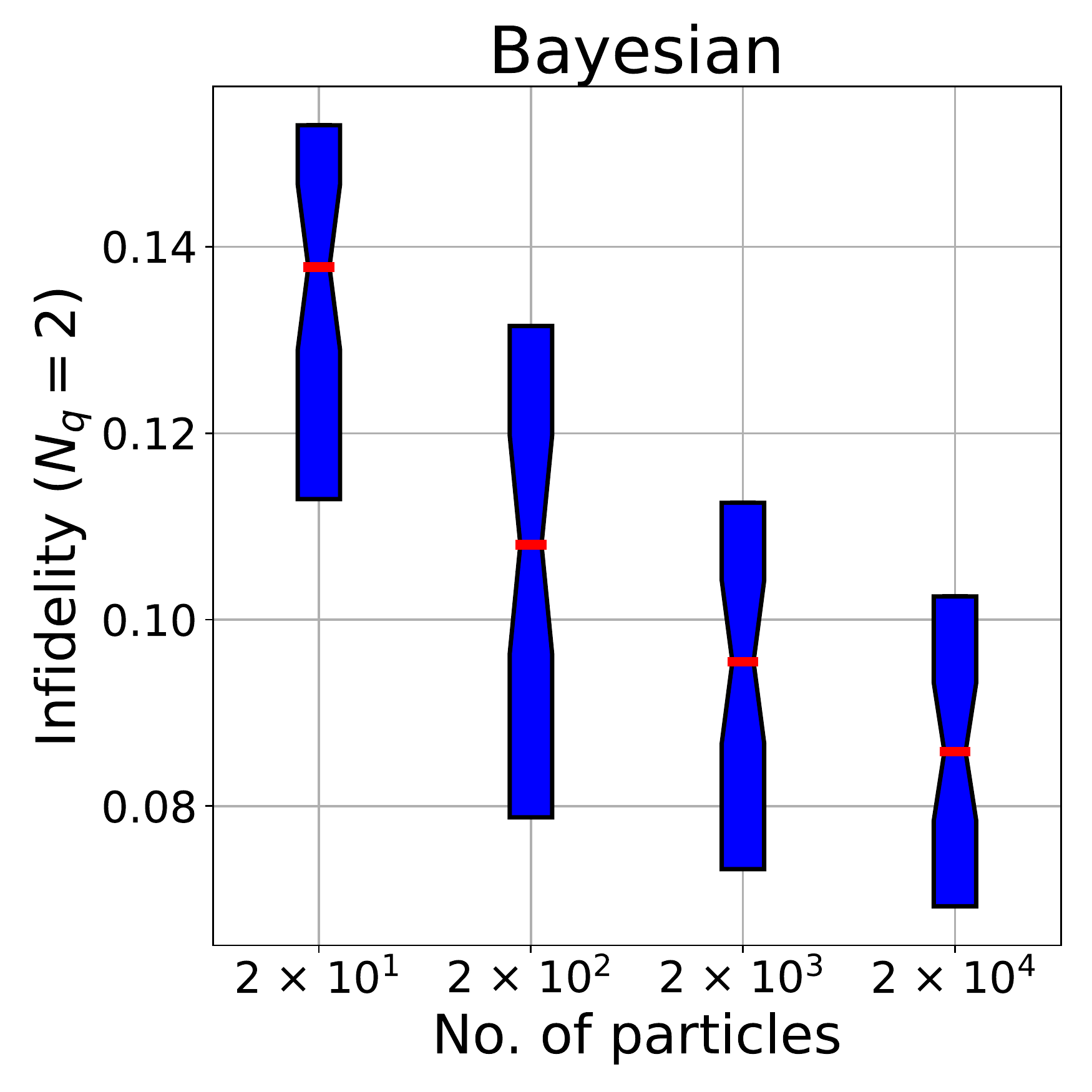}
\caption{Bayesian reconstruction of random two-qubit state against particle filter sample sizes, averaged over 50 trials. In each case the number of measurements was taken to be $N = 100$ due to the computation time taken for higher sample sizes.}
\label{Bay2Q}
\end{figure}

\begin{figure}[t!]
\center
\includegraphics[width=.75\linewidth]{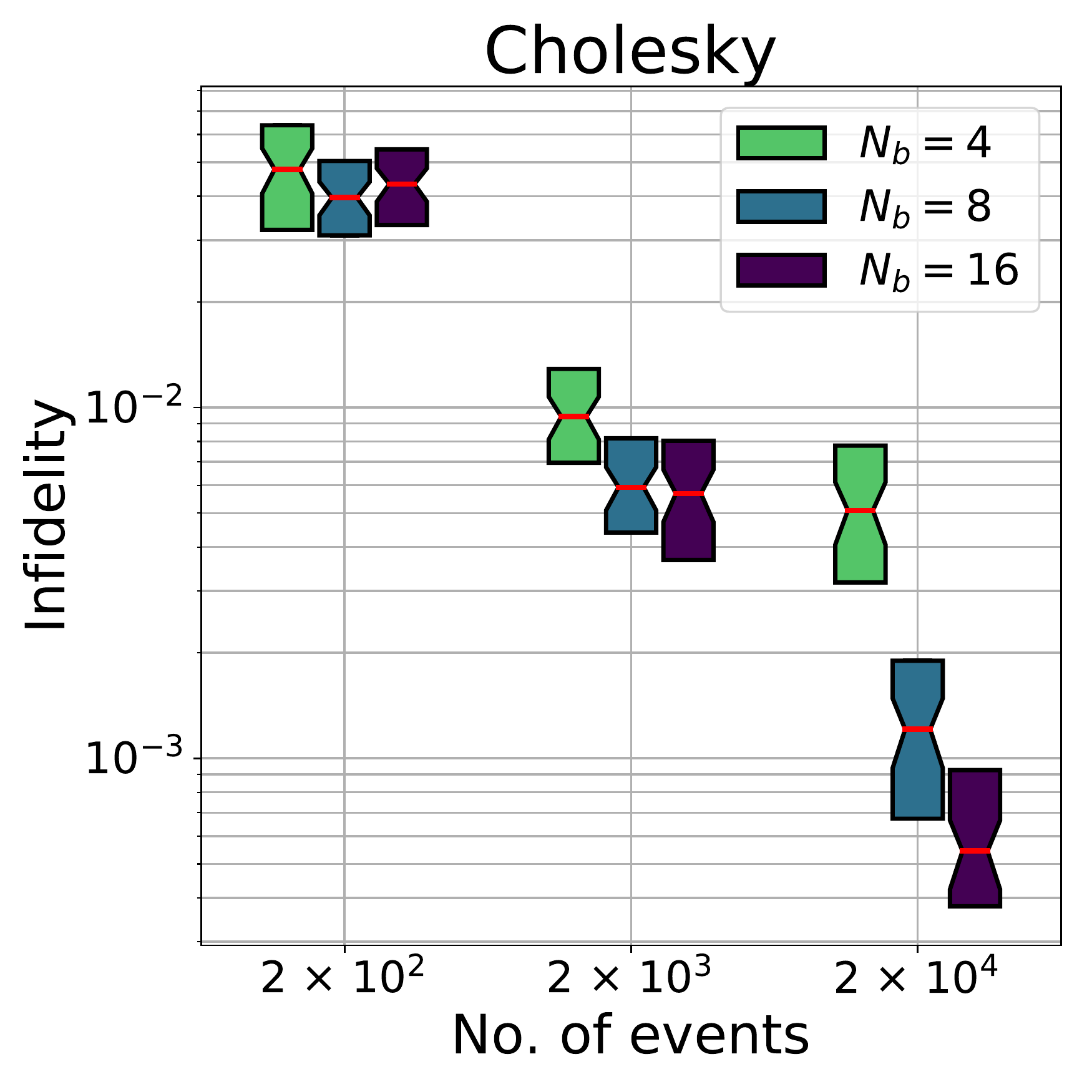} 
\caption{Full rank two-qubit reconstruction infidelity using the Cholesky method results for 4,8 and 16 segments, with increasing number of measurements and averaged over 50 trials.}
\label{2Q}
\end{figure}

Fig~\ref{Bay2Q} and Fig~\ref{2Q} show the effect of particle filter sample sizes on a Bayesian two-qubit reconstruction, and the performance of the Cholesky method for a two-qubit reconstruction, respectively.

\section{Pseudocodes}
\label{app:pseudo}

In this section we present the pseudocodes for the PGDB algorithm, and some subroutines used for the Bayesian approach taken from \cite{Granade2012}. 

\begin{algorithm}[H]
\caption{PGDB}
\label{PGDB}
\begin{algorithmic}[1]
\State $k=0$, $\mu_{k=0} = 1$
\State Initial estimate $\rho_{k=0} \in \mathcal{S}$. 
\State Given $\delta = 10^{-4}$, $\gamma = 10^{-3}$, $\mu_{min} = 10^{-4}$, $\mu_{max} = 10^4$
\While{$\sum^{20}_{i=1} |\mathcal{C}(\rho_i) - \mathcal{C}(\rho_{i-1})| > \delta$}
    \State Calculate probability estimates 
    \State Calculate log likelihood $\mathcal{C}(\rho_k) = -\sum_i n_i \mathrm{log}(p_i)$
    \State Calculate gradient $\nabla \mathcal{C}(\rho_k)= -\sum_i (n_i / p_i) \Ket{\phi_i}\Bra{\phi_i}$
    \State $D_k = \mathcal{P}_S (\rho_k - \mu^{-1}_k \nabla \mathcal{C}) - \rho_k$
    \State $\tilde{\mathcal{C}}(\rho_k) = \mathcal{C}(\rho_k) + \gamma \mathrm{Tr}[D_k \mathcal{C}(\rho_k)]$
    \State Initialise line search parameter $\alpha = 1$
    \While{$\mathcal{C}(\rho_k + \alpha D_k) > \tilde{\mathcal{C}}(\rho_k)$}
        \State $\alpha = \alpha / 2$
    	\State $\tilde{\mathcal{C}}(\rho_k) =  \mathcal{C}(\rho_k) + \gamma \alpha \mathrm{Tr}[D_k \mathcal{C}(\rho_k)]$
    \EndWhile
    \State $\rho_{k+1} = \rho_k + \alpha D_k$
    \State $\mu_{k+1} = \mathrm{min} \{ \mathrm{max} \{ \frac{\langle \rho_k - \rho_{k-1}, \nabla \mathcal{C}(\rho_k) - \nabla \mathcal{C}(\rho_{k-1}) \rangle}{\| \rho_k - \rho_{k-1} \|^2}, \mu_{min} \}, \mu_{max} \} $ \Comment Update scale factor for step in gradient direction \cite{Goncalves2016}
    \State $k = k+1$
\EndWhile
\State \Return $\rho_{end} = \mathcal{P}_S (\rho_{k+1})$
\end{algorithmic}
\end{algorithm}

\begin{algorithm}[H]
\caption{SMC update algorithm}
\label{SMCupdate}
\begin{algorithmic}[1]
\State Initial distribution for particle positions $\{ \mathbf{x}_j \}$ and weights $\{ w_j \}$ \Comment Chosen to be both uniform
\For{$i \in \mathrm{range}(N)$}
	\State New datum $D_i = \{ \alpha_i, \mu_i \}$ is measured
	\For{$j \in \mathrm{range}(n_{part})$}
		\State $w_j = w_j P(\{ \alpha_i, \mu_i \}|\mathbf{x}_j)$
	\EndFor
	\State Renormalise $\{ w_j \}$
\EndFor
\end{algorithmic}
\end{algorithm}

\begin{algorithm}[H]
\caption{SMC resampling algorithm}
\label{SMCresamp}
\begin{algorithmic}[1]
\Function{resample}{$\{ \mathbf{x}_j \}, \{ w_j \}$, a}
	\State $\boldsymbol{\mu} = \mathrm{MEAN}(\{ \mathbf{x}_j \}, \{ w_j \})$ \Comment Weighted mean of $\{ \mathbf{x}_j \}$
	\State $h = \sqrt{1-a^2}$
	\State $\boldsymbol{\Sigma} = \mathrm{COV}(\{ \mathbf{x}_j \}, \{ w_j \})$ \Comment Find covariance
	\For{$i \in \mathrm{range}(n_{part})$}
		\State Select particle $\mathbf{x}_j$ with probability $w_j$
		\State $\boldsymbol{\mu}_i = a \mathbf{x}_j + (1-a) \boldsymbol{\mu}$ \Comment Mean for new particle location
		\State Pick $\mathbf{x}'_i$ randomly from $\mathcal{N}(\boldsymbol{\mu}_i, \boldsymbol{\Sigma})$ \Comment Draw new, shifted, particle
		\State $w'_i = n^{-1}_{part}$ \Comment Reset weights to uniform 
		
	\EndFor \\
	\Return $\{ \mathbf{x}'_j \}, \{ w'_j \}$
\EndFunction
\end{algorithmic}
\end{algorithm}

Algorithm \ref{SMCresamp} can then be added to Algorithm \ref{SMCupdate}, conditioned on the value of the effective sample size $n_{eff} = 1/ \sum_i w_i$. If $n_{eff}$ is less than some threshold value (taken to be 0.5 \cite{Granade2012}), then the distribution is resampled. Details on the MEAN and COV functions can be found in Ref.~\citen{Granade2012}

\section{Alternative summary of PGDB performance, with error bars}
\label{app:PGDBalt}
\begin{figure}[h!]
\center
\includegraphics[width=.9\linewidth]{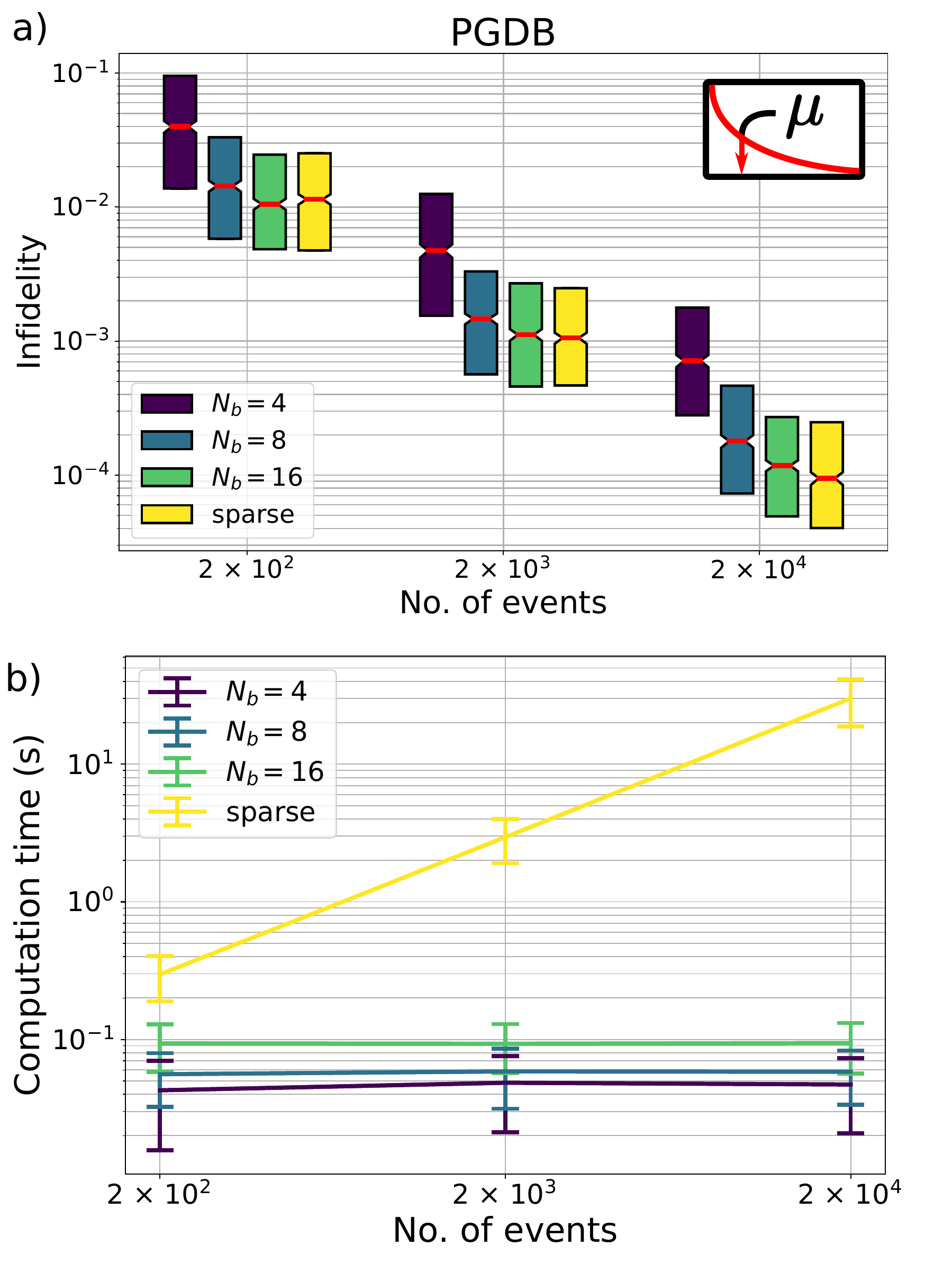}
\caption{a) Full rank, single qubit reconstruction using gradient descent, averaged over 1000 trials. The random phases were sampled from an exponential distribution with  $\mu = \pi/8$. As expected the coarse grained approach returns slightly higher infidelities. b) Algorithm running times for the unsorted, and coarse grained approaches. The unsorted approach scales linearly with number of measurement repetitions. The coarse grained approaches, within the standard deviation, do not scale with increased repetitions as the number of projective operators used for reconstruction is the same for all repetition numbers. }
\label{pgdb_coarse}
\end{figure}

In this section we re-state the performance of PGDB for the exponential distribution, but in an alternative format with error bars: see Fig.~\ref{pgdb_coarse}.

 \clearpage

\bibliography{retrocite}

\end{document}